\documentclass[a4paper,11pt]{article}
\usepackage{pos}
\newcommand{\beq}{\begin{equation}}
\newcommand{\eeq}{\end{equation}}

\newcommand{\GeV}{\;{\mathrm{GeV}}}
\newcommand{\GeVc}{\;{\mathrm{GeV/c^2}}}

\title{Axial and Vector Structure Functions for  Lepton-Nucleon Scattering, NuFact 2021 Update}

\author*[a]{Arie Bodek}
\author[b]{Un-ki Yang}
\author[b]{Yang Xu}

\affiliation[a]{university of Rochester, Department of Physics and Astronomy,\\
  Rochester, NY 14627-0171, USA}

\affiliation[b]{Department of Physics and Astronomy,\\
Seoul National University, Seoul 151-747, Korea}

\emailAdd{Arie Bodek <bodek@pas.rochester.edu>}
\emailAdd{Un-ki Yang <ukyang@snu.ac.kr>}
\emailAdd{Yang Xu <yxu100@ur.rochester.edu>}

\abstract{We report on an update (2021) of a phenomenological model for inelastic neutrino- and electron-nucleon  scattering cross sections using effective leading order parton distribution functions with a new scaling variable $\xi_w$.  Non-perturbative effects  are  well described using the  $\xi_w$ scaling variable in combination with multiplicative $K$ factors at low $Q^2$. The model describes all inelastic charged-lepton-nucleon scattering data (HERA/NMC/BCDMS/SLAC/JLab) ranging from very high $Q^2$  to very low $Q^2$ and down to the $Q^2=0$ photo-production region. The model has been developed to be  used in analyses of  neutrino oscillation experiments  in the few$\GeV$ region.  The 2021 update  accounts for the difference between axial and vector structure functions which brings it into much better agreement with  neutrino-nucleon total cross section measurements. The model has been developed primarily for  hadronic final state masses  $W$ above 1.8 GeV. However with additional parameters  the model  also describes the $average$ neutrino cross sections in the resonance region down to $W$=1.4 GeV.}
\FullConference{%
  *** The 22nd International Workshop on Neutrinos from Accelerators (NuFact2021) ***\\
  *** 6-11 Sep 2021 ***\\
  *** Cagliari, Italy ***}

\begin{document}
\maketitle

\section{Introduction}

In the few $\GeV$ region there are contributions from several kinds of lepton-nucleon interaction processes as defined by the final state invariant mass $W$ and square of the momentum transfer $Q^2$.   These  include quasi-elastic reactions ($W<1.07~\GeVc$),  the  $\Delta(1232)$  region  ($1.1<W<1.4~\GeVc$ ),  higher mass resonances  ($1.4<W<2.0~\GeVc$), and the  inelastic continuum region ($W> 2.0~\GeVc$).  At low momentum transfer the inelastic continuum is sometimes referred to as "shallow inelastic", and at high momentum transfer it is referred to as "deep inelastic".  It is quite challenging to disentangle each of those contributions  separately, and in particular  the  contribution of resonance production and the  inelastic scattering continuum.  At low $Q^2$  there are large non-perturbative  contributions to the inelastic cross section. These include kinematic target mass corrections,  dynamic higher twist effects,  higher order Quantum Chromodynamic (QCD) terms, and nuclear effects in nuclear targets.   

In the Bodek-Yang model  we focus on the inelastic part of the cross sections above the region of the $\Delta$(1232) resonance (i.e. the  higher mass resonances, and the inelastic continuum).  The model is duality based model of neutrino interactions using effective leading order parton distribution functions (PDFs). Earlier  versions of the model\cite{nuint01-2,nuint04} have been incorporated into several Monte Carlo generators of neutrino interactions including NEUT,   NEUGEN,   NUANCE and GENIE.  The current version of GENIE  is using the NUINT04\cite{nuint04} version of the model. These early versions  assume that the axial structure functions are the same as the vector structure functions.   The model is based on parameters extracted from electron scattering data. The leading order GRV98 PDFs are used with modifications that include  a new scaling variable ($\xi_w$) to account for deviations from Bjorken scaling at lower values of $Q^2$ and low $Q^2$ K factors that extend the validity of the model down  the $Q^2$=0 photo-production limit. Figure 1 shows a comparison of electron scattering structure functions to predictions of the GRV98 PDFs, with and without our modifications.

In this conference report we present a short a summary of the  results of a  2021 update  (details in ref.\cite{update} in which we  further refine the model and $also$ account for the difference  between axial and vector structure functions at low values of $Q^2$. We refer to the version of the model which assumes that vector and axial structure functions are the same as "Type I". The "Type I" version should be used to model electron and muon scattering.  We refer to the updated version of the model that accounts for the difference in vector and axial structure functions as "Type II".  "Type II" model should be used to model neutrino scattering.   

Figure   1 shows the  Bodek-Yang model with  effective LO PDF model  compared to charged-lepton  ${\cal F}_2$ experimental data (SLAC, BCDMS, NMC). Left: ${\cal F}_2$ proton. Right ${\cal F}_2$ deuteron (per nucleon). The solid lines are the model, and the  dashed lines are predictions of the GRV98  PDFs without our low $Q^2$ modifications\
\begin{figure}
\includegraphics[width=3.in,height=3.5in]{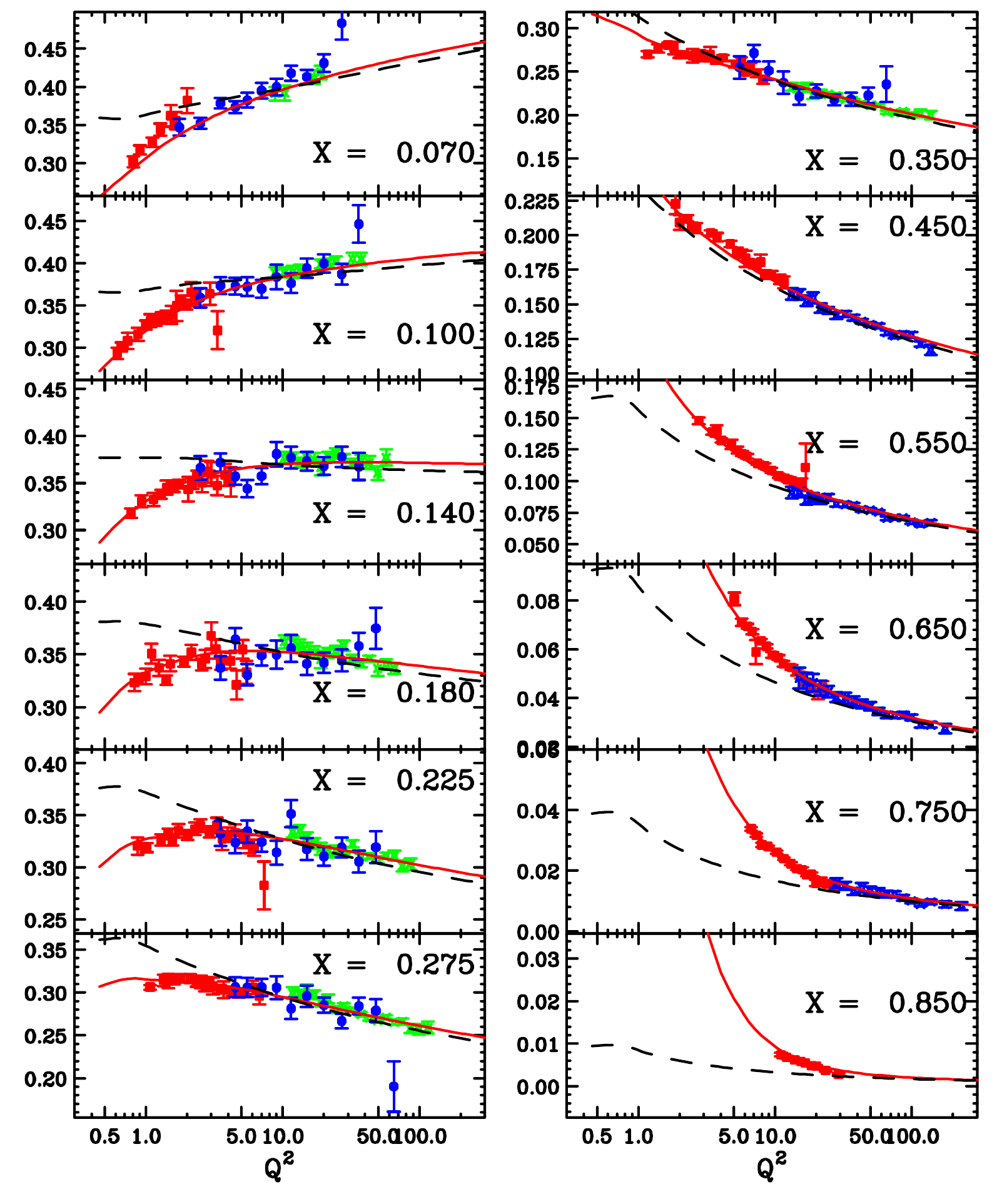}
\includegraphics[width=3.in,height=3.5in]{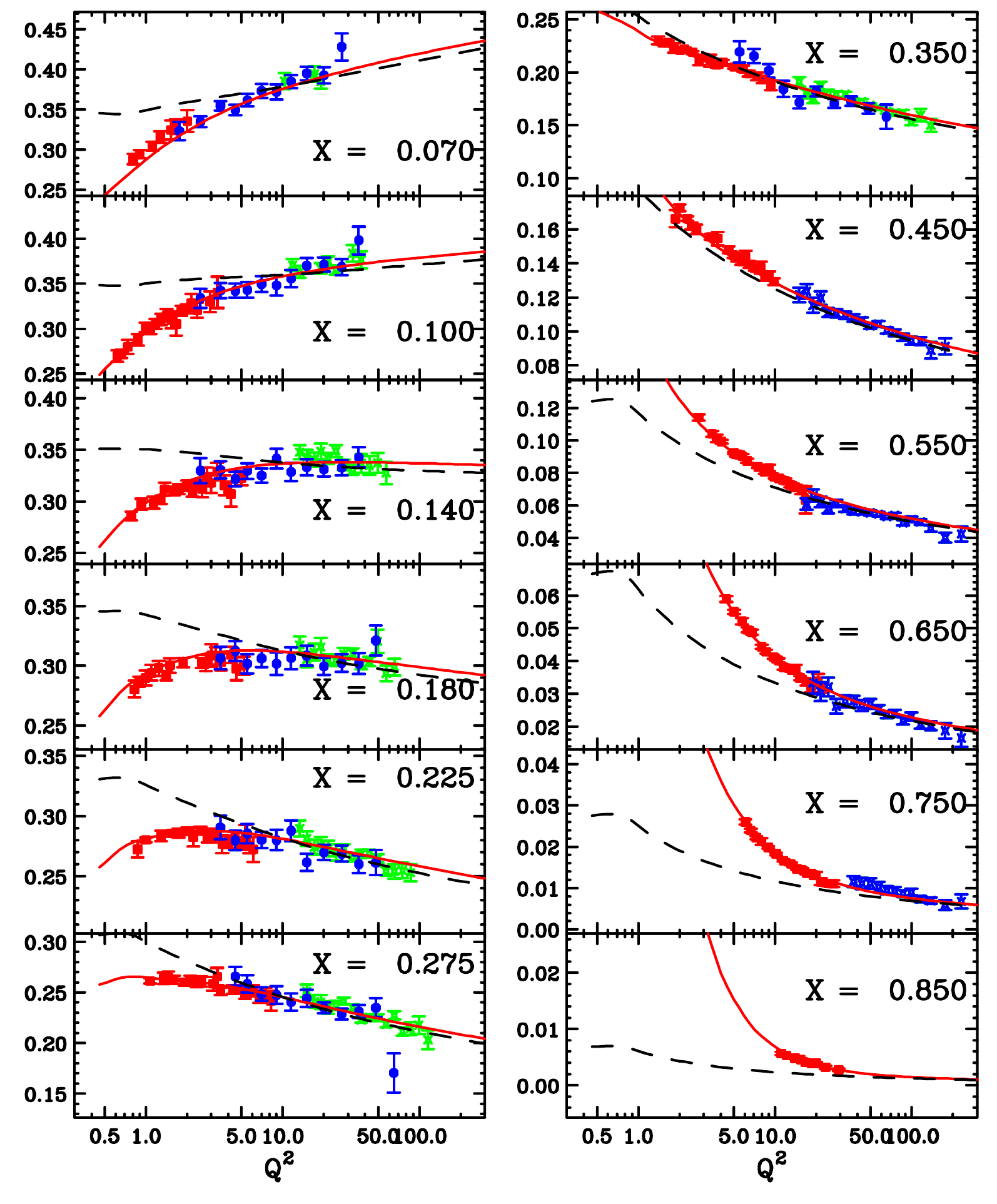}
\caption{The Bodek-Yang model with  effective LO PDF model  compared to charged-lepton  ${\cal F}_2$ experimental data (SLAC, BCDMS, NMC). Left: ${\cal F}_2$ proton. Right ${\cal F}_2$ deuteron (per nucleon). The solid lines are the model, and the  dashed lines are predictions of the GRV98  PDFs without our low $Q^2$ modifications.}
\label{fig:f2fit_highx}
\end{figure}

Figure 2 shows the model  predictions (per nucleon) for neutrino total cross sections on an  isoscalar  target  compared to measurements. $\sigma_{\nu}$/E data as a function of energy are shown on the left and  $\sigma_{\bar \nu}$/E are shown on the right.
The green points are MINOS $\sigma_{\nu}$/E. data, the blue points are NOMAD,  and the yellow crosses are  BNL82.   The MINERvA and T2K data are shown in purple and brown, respectively.
The Gargamelle  and ArgoNeut measurements of  $\sigma_{\bar \nu}$/E per nucleon are identified on the figure.  There is good agreement of the the Type II (A$>$V)  model predictions with neutrino and antineutrino total cross section measurements.
 \begin{figure}
\includegraphics[width=2.9 in,height=2.1in]{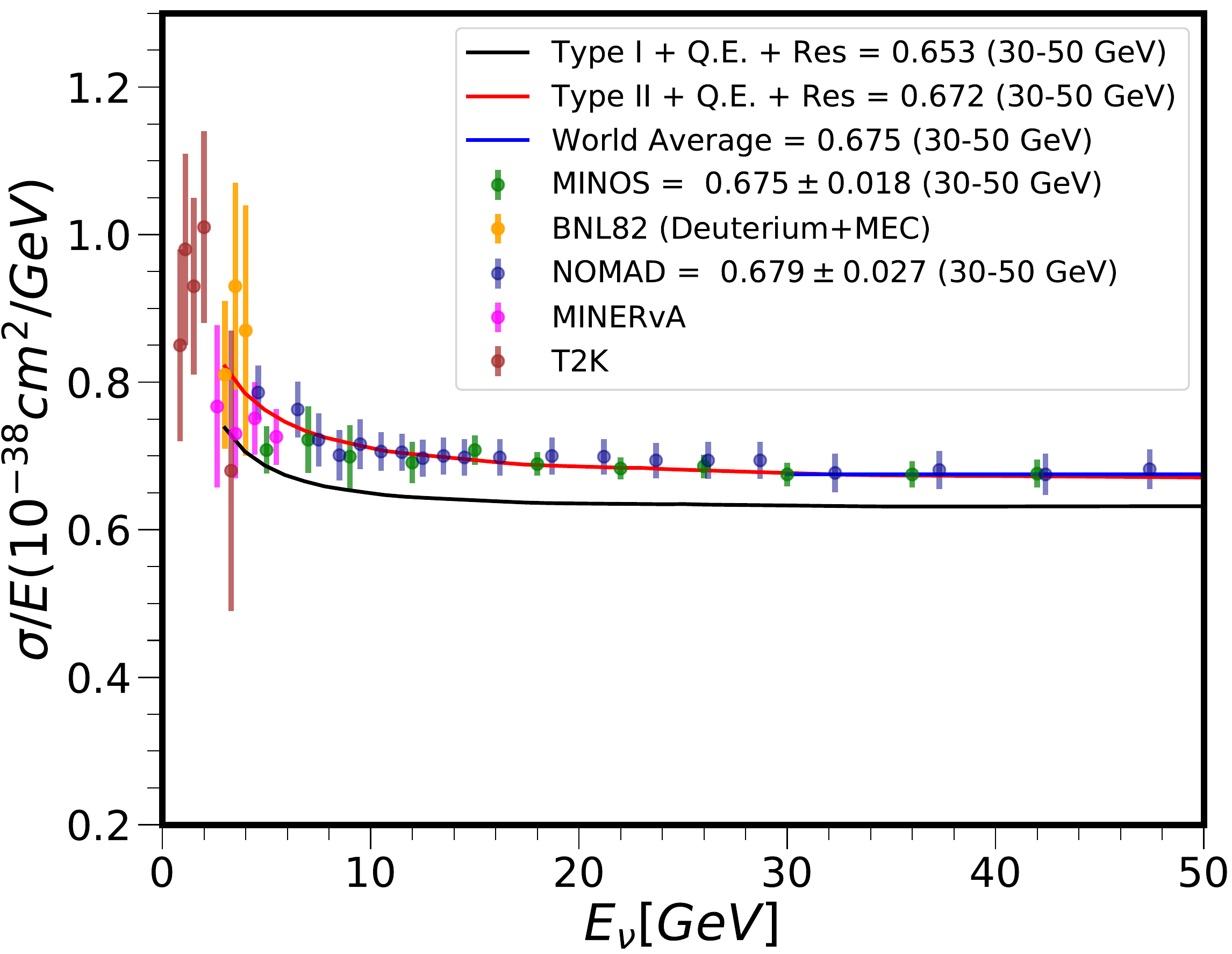}
\includegraphics[width=2.9 in,height=2.1in]{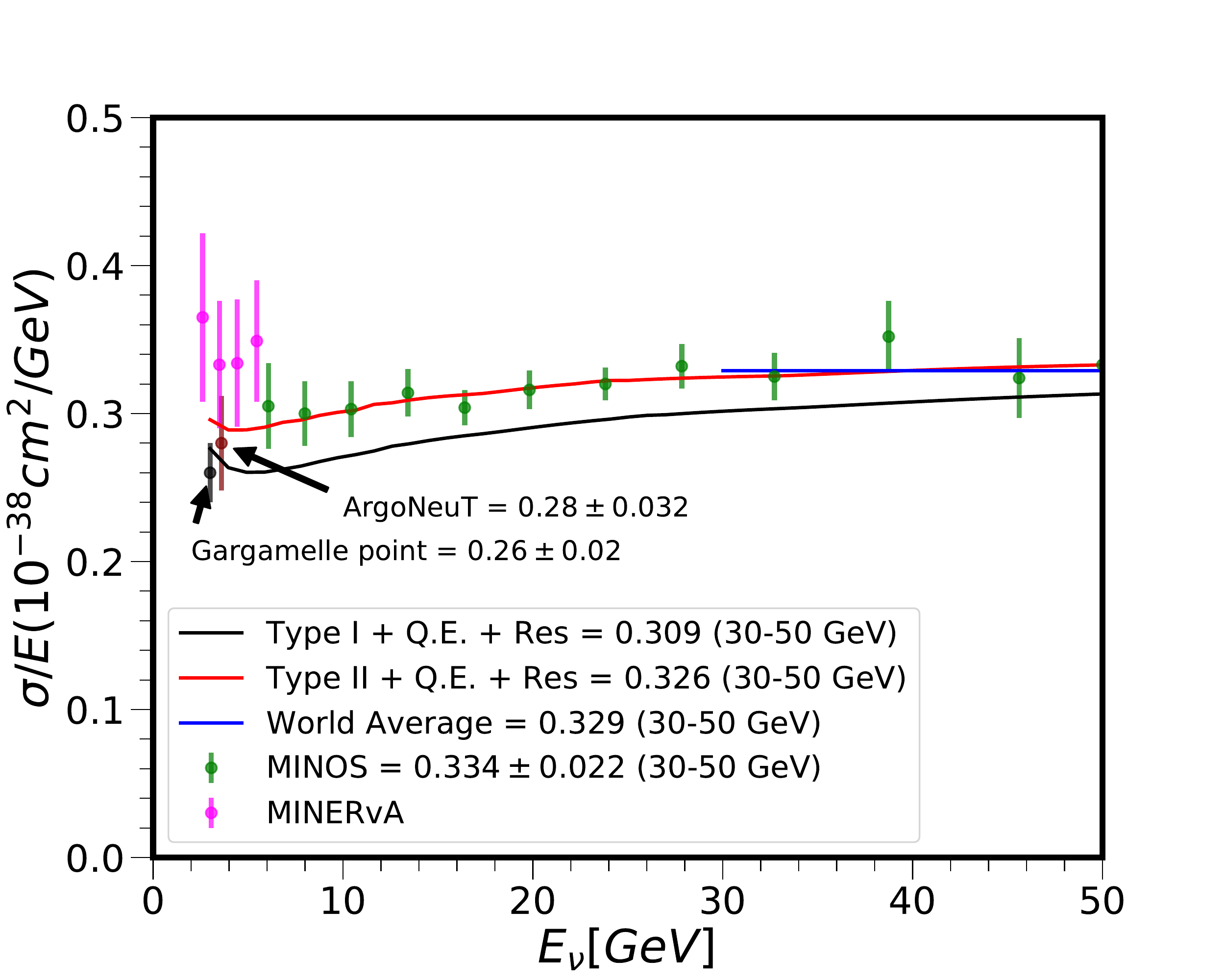}
\caption {Model  predictions (per nucleon) for neutrino total cross sections on an  isoscalar  target  compared to measurements. $\sigma_{\nu}$/E data as a function of energy is shown on the left and  $\sigma_{\bar \nu}$/E data are shown on the right.
The green points are MINOS data, the blue points are NOMAD and the yellow crosses are  BNL82.  The MINERvA and T2K data are shown in purple and brown, respectively.
The Gargamelle  and ArgoNeut measurements of  $\sigma_{\bar \nu}$/E per nucleon are identified on the figure. There is good agreement of the Type II (A$>$V)  model predfictions with neutrino and antineutrino total cross section measurements.
}
\label{fig:neutrinoD12}
\end{figure}

Figure 3  shows the ratio of  charged-current neutrino and antineutrino 
differential cross sections $d^2\sigma/dxdy$ on lead from CHORUS  (blue points)
 to the  Type II (A$>$V) default model predictions. The ratios are shown for energies of 15 and  25  $GeV$.    On the left side of each panel  we show the comparison for neutrino cross sections and on the right side we show the comparisons for antineutrinos.   The black line is the ratio of the predictions of the  Type I (A=V) model for which the  axial structure functions are set equal to the vector structure functions,  to the predictions of the  Type II (A$>$V)  default model. 
 The differential cross section data favor the  Type II (A$>$V)  model. 
  \begin{figure}
\includegraphics[width=3.0 in,height=3.3in]{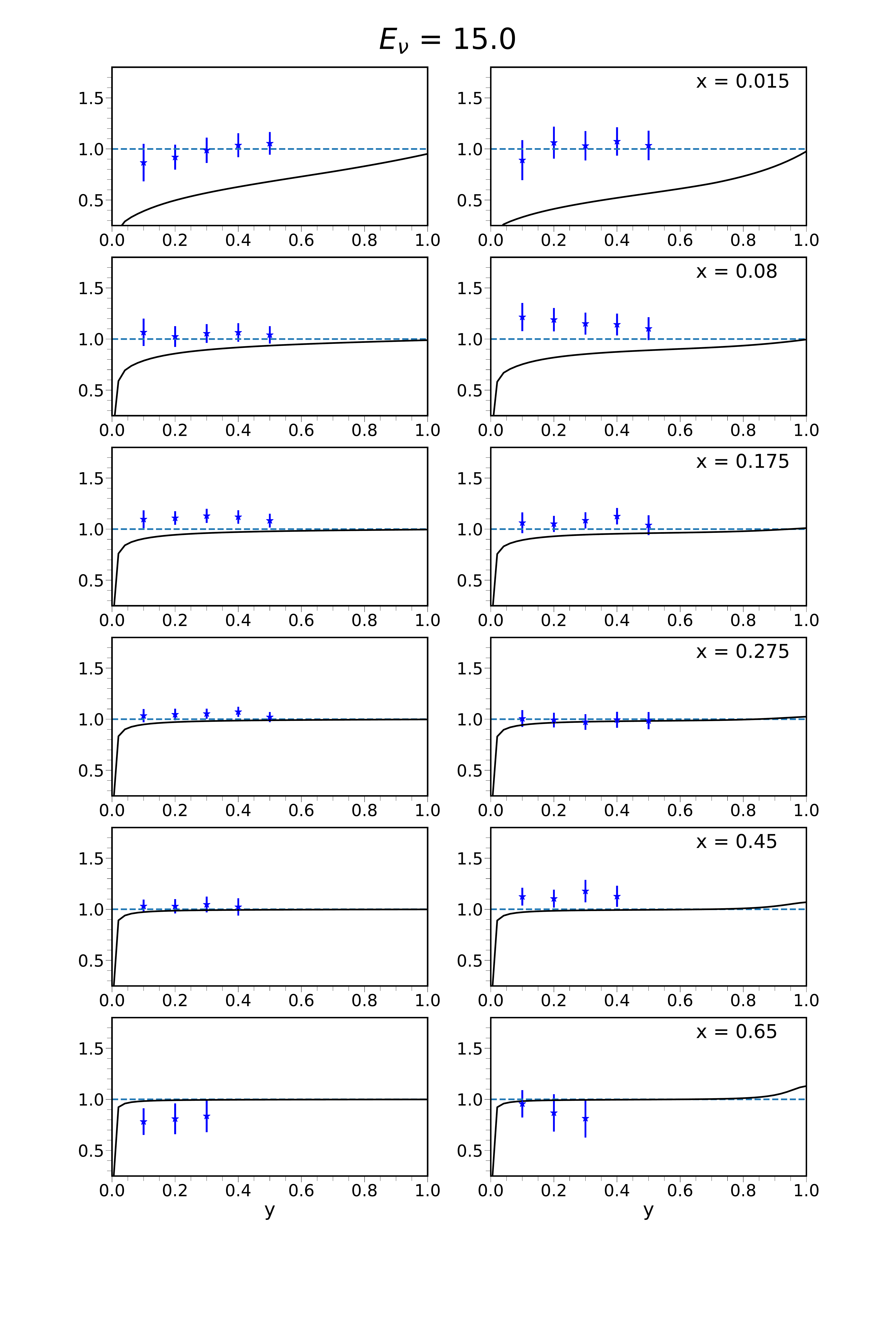}
\includegraphics[width=3.0 in,height=3.3 in]{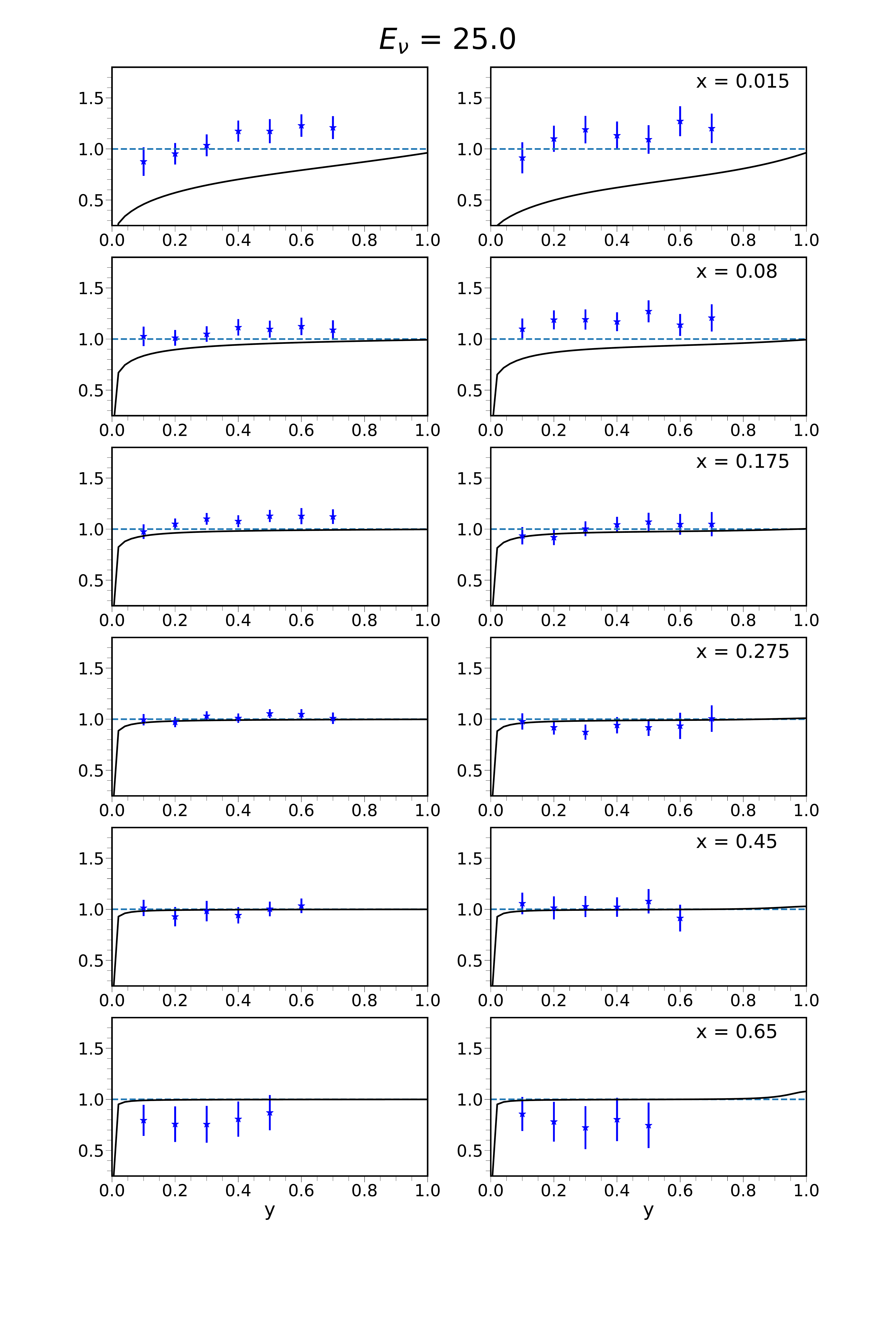}
\caption { The ratio of  charged-current neutrino and antineutrino 
differential cross sections $d^2\sigma/dxdy$ on lead from CHORUS  (blue points) 
to the  Type II (A$>$V) default model  predictions. The ratios are shown for  energies of 15 and 25 $GeV$.   On the left side of each panel we show the comparison for neutrino cross sections and on the right side we show the comparisons for antineutrinos.   The black line is the ratio of the predictions of the  Type I (A=V) model for which the  axial structure functions are set equal to the vector structure functions,  to the predictions of the  Type II (A$>$V)  default model.  The differential cross section data favor the  Type II (A$>$V)  model. }
\label{fig:neutrinoD3} 
\end{figure}
In conclusions, the predictions of the 2021 update of the Bodek-Yang model are in excellent agreement with both neutrino and charged lepton cross section measurements on  nuclear targets

\end{document}